\documentclass[conference]{IEEEtran}
\IEEEoverridecommandlockouts

\usepackage{cite,comment}
\usepackage{amsmath,amssymb,amsfonts}
\usepackage{algorithmic}
\usepackage{algorithm}  
\usepackage{graphicx}
\usepackage{textcomp}
\usepackage{xcolor}
\usepackage{url}
\usepackage{bm}
\usepackage{booktabs}
\usepackage{subcaption}
\def\BibTeX{{\rm B\kern-.05em{\sc i\kern-.025em b}\kern-.08em
    T\kern-.1667em\lower.7ex\hbox{E}\kern-.125emX}}
\begin{document}

\title{Are Bus-Mounted Edge Servers Feasible?
}
\author{
	\IEEEauthorblockN{\centering
		Xuezhi Li\IEEEauthorrefmark{1}, 
		Jiancong He\IEEEauthorrefmark{1}, 
		Ming Xie\IEEEauthorrefmark{1}, 
        Xuyang Chen\IEEEauthorrefmark{1},
		Le Chang\IEEEauthorrefmark{1}, 
        Li Jiang\IEEEauthorrefmark{1}\textsuperscript{1},
        and Gui Gui\IEEEauthorrefmark{2}\IEEEauthorrefmark{3}} 
	\IEEEauthorblockA{\IEEEauthorrefmark{1}School of Automation, Guangdong University of Technology, Guangzhou 510006, China}
	\IEEEauthorblockA{\IEEEauthorrefmark{2}School of Automation, Central South University, Changsha 410083, China}
	\IEEEauthorblockA{\IEEEauthorrefmark{3}Hunan Provincial Key Laboratory of Energy Saving Control and Safety Monitoring for Rail Transportation,\\ Changsha 410083, China}
	\thanks{
	$^1$Corresponding author: jiangli@gdut.edu.cn}
}

\maketitle

\begin{abstract}
Placement of edge servers is the prerequisite of provisioning edge computing services for Internet of Vehicles (IoV). Fixed-site edge servers at Road Side Units (RSUs) or base stations are able to offer basic service coverage for end users, i.e., vehicles on road. However, the server locations and capacity are fixed after deployment, rendering their inefficiency in handling spationtemporal user dynamics. Mobile servers such as buses, on the other hand, have the potential of adding computation elasticity to such system. To this end, this paper studies the feasibility of bus-mounted edge servers based on real traces.
First, we investigate the coverage of the buses and base stations using the Shanghai bus/taxi/Telecom datasets, which shows a great potential of bus-based edge servers as they cover a great portion of geographic area and demand points. 
Next, we build a mathematical model and design a simple greedy heuristic algorithm to select a limited number of buses that maximizes the coverage of demand points, i.e., with a limited purchase budget. 
We perform trace-driven simulations to verify the performance of the proposed bus selection algorithm. The results show that our approach effectively handles the dynamic user demand under realistic constraints such as server capacity and purchase quantity. Thus, we claim:  bus-mounted edge servers for vehicular networks in urban areas are feasible, beneficial, and valuable.
\end{abstract}


\section{Introduction}
Recently, edge computing has been recognized as one of the key enabling technologies for Internet of Vehicles (IoV) and autonomous driving, which fundamentally reforms the traditional data processing paradigm. 
Edge servers are physically closer to end users, enabling low latency, localized processing, and secured service for interactive applications~\cite{mao2017survey,ranaweera2021survey}. 

The placement of edge servers has attracted significant attention, as placement strategies directly impact the overall edge computing performance. Existing works have proposed deploying edge servers on roadside units (RSUs) or base stations, allowing data to be processed locally on road which
effectively improves the task completion ratio and service continuity~\cite{fan2025vehicular}. However, as these fixed-site servers have fixed computation capacity and no mobility, it is difficult for them to handle spatiotemporally dynamic user demand, especially for fast-moving vehicles.
To add computation elasticity, deploying edge servers on mobile units such as Unmanned Aerial Vehicles (UAVs) or service vehicles have been proposed~\cite{zhou2020mobile,xie2025multi}.  
The superior mobility of such mobile units allows highly flexible deployment strategies that fit the rapid movement of vehicles. However, UAV-mounted edge servers suffer from limited payload and battery capacity, etc., which restrict its practical application.

Up to now, most of the existing works have overlooked an ideal type of edge server carriers: public buses. In fact, buses have many inherent advantages in carrying edge servers. First, buses are readily available resources that require no additional land acquisition or road space occupation. Second, buses have sufficient space to accommodate powerful servers and high-capacity batteries, which are crucial for provisioning stable edge computing services. Third, 
buses are close to end users, i.e., vehicles on road, which allows fast and direct V2V communication. Fourth, there are a great number of buses operating on road everyday, offering sufficient coverage area and selection space. Last but not the least important, buses and end users usually share the same  mobility pattern, i.e., more buses are dispatched during rush hours, which facilitates tackling the demand hotspots with enough buses nearby.    

Therefore, this paper studies the feasibility of deploying edge servers on public buses. We aim to answer the following questions. Are bus-mounted edge servers feasible in urban vehicular networks? How much benefit can we gain from deploying edge servers on buses? How to select a given number of buses to satisfy a maximum portion of user demand? To answer these questions, our main work and contributions in this paper are summarized as follows.
\begin{itemize}
    \item We conduct a trace study to show the potential of public buses being edge servers in Shanghai urban area, i.e., sufficient coverage for handling the spatiotemporally distributed computation requests of end users. 
    \item We build a mathematical model of maximizing the covered user demand with a fixed number of buses, i.e., under a limited budget,  and 
    propose a heuristic algorithm to select most suitable buses from thousands of candidates.
    \item We run trace-driven simulations to verify the advantage of bus-mounted edge servers and the proposed bus selection algorithm. The results show that our method can optimize the spatiotemporal coverage of bus-mounted edge servers,  and fill the service gap of base stations.
\end{itemize}

The remainder of this paper is organized as follows. Section~\ref{sec:rela} lists the related works. Section~\ref{sec:trace} describes our trace study on the coverage of buses, validating their potential of being edge servers. Section~\ref{sec:selection} presents our system model and bus selection algorithm, and Section~\ref{sec:eval} evaluates its performance. Section~\ref{sec:conclu} concludes the paper.

\section{Related Works}\label{sec:rela}

Hosono et al. proposed a collaborative load-balancing mechanism for edge servers based on Lane Section ID, which effectively mitigates local data processing overload caused by base station signal coverage deviation~\cite{hosono2021implementation}. 
Feng et al. developed a joint offloading and resource allocation strategy for C-V2X vehicular networks, 
and designed a greedy offloading algorithm (PC5-GO) that dynamically coordinates the computation resources between RSUs and neighboring cooperative vehicles (CVs), achieving proactive load balancing and latency optimization under dual-interface constraints~~\cite{feng2022joint}.

Regarding the feasibility of deploying edge servers on public buses, Liu et al. conducted a comprehensive survey on Vehicular Edge Computing (VEC), focusing on the potential of utilizing different vehicle types as edge servers, and analyzing their applications and challenges in collaborative offloading, distributed caching, and dynamic resource management~\cite{liu2021vehicular}. Lin et al. proposed an Edge-Computing-Based Public Vehicle System (ECPV), where RSUs act as edge devices to enable real-time vehicle–passenger matching, providing practical validation for buses as edge computing nodes~\cite{lin2020edge}. 
Carvalho et al. reviewed edge computing applications in vehicular networks, highlighting both the advantages (fixed routes, stable power supply) and challenges (mobility, resource constraints) of buses as edge nodes, and providing a theoretical and practical framework for deploying edge resources in intelligent transportation systems~\cite{carvalho2021edge}. 
Zhou et al. explored Edge-of-Things (EoT) applications in smart cities, emphasizing the architectural advantages of RSUs and vehicles as edge nodes, thereby offering theoretical support and empirical validation for deploying buses and other transport vehicles as mobile edge servers~\cite{zhou2022emerging}.
Zhang et al. proposed bus-based mobile edge server deployment, using deep Q-learning to optimize server selection, aiming to mitigate frequent handovers and high latency caused by vehicle mobility~\cite{zhang2020moving}. 
Similarly, Deng et al. proposed bus-mounted edge servers and employed an improved Dung Beetle Optimization (DBO) algorithm for computation offloading, aiming to reduce latency, energy consumption, and cost~\cite{deng2024computation}. 

However, placement design of bus-mounted edge servers is still in its infancy stage, with very limited work published. To the best of our knowledge, this paper is among the first works that study the feasibility and placement strategies of bus-mounted edge servers based on real trace data.

\section{The Coverage of Buses in Shanghai Urban Area: A Trace Study}\label{sec:trace}
In this section, we investigate the coverage of public buses in comparison with base stations in Shanghai urban area through a trace study, to show their potential of being edge servers.
\subsection{Datasets}
We use three datasets to evaluate the coverage and placement strategies of the buses: the Shanghai Telecom dataset~\cite{li2021profit,guo2020user,wang2019delay}, the Shanghai bus dataset~\cite{qiangsheng2018}, and the Shanghai taxi dataset~\cite{SUVnet_Trace_Data}. 

\textbf{Shanghai Telecom Dataset}:
This dataset is released by Shanghai Telecom, which contains over $7.2$ million records collected in six months, documenting the Internet accesses by $9,481$ mobile phones through $3,233$ base stations. We extract the geographic coordinates of the base stations from the trace to derive the coverage of fixed edge servers, assuming that they are deployed at some of these base stations.

\textbf{Shanghai Bus Dataset}:
This dataset includes over $5$ million GPS records from over $2,100$ buses, during the period from February 19 to February 23 in 2007. 
In this dataset, we extract the longitude, latitude, time, and ID fields of the buses to construct their trajectories and derive the coverage. We treat these buses as potential edge servers which are able to handle computation requests generated by end users.

\textbf{Shanghai Taxi Dataset}:
We use the Shanghai Qiangsheng Taxi dataset which are publicly available online. This dataset records the taxi trajectories in Shanghai on April 1, 2018, and includes approximately $85$ million records from $11,936$ vehicles. We extract the longitude, latitude, time, and ID fields of the taxis from the trace and consider them as the end users that generate computation requests, i.e., demand points.

The datasets span over a wide range of the Shanghai city, but we select the center of the map and partition it into grid cells with a side length of $200$ meters. The selected region is a square area that covers $72\times72$ grids, containing $934$ base stations, $2,143$ buses, and over $26$ million taxi GPS records. 
The high density of the base stations, buses, and taxis in this area allows us to build a thorough understanding of the coverage of buses and base stations from the trace data. We divide the records in the trace into discrete time slots (e.g., 30 seconds each). For each time slot, we retrieve the locations of the base staions, buses, and taxis, and evaluate how many grids or taxis (demand points) can be covered by the base stations or buses, given a service radius $R$ of these potential servers.

\subsection{The Coverage of Potential Edge Servers}
 We introduce a family of metrics to quantify their coverage as follows.
\begin{figure}[!htp] 
  \centering
  \includegraphics[width=0.85\linewidth]{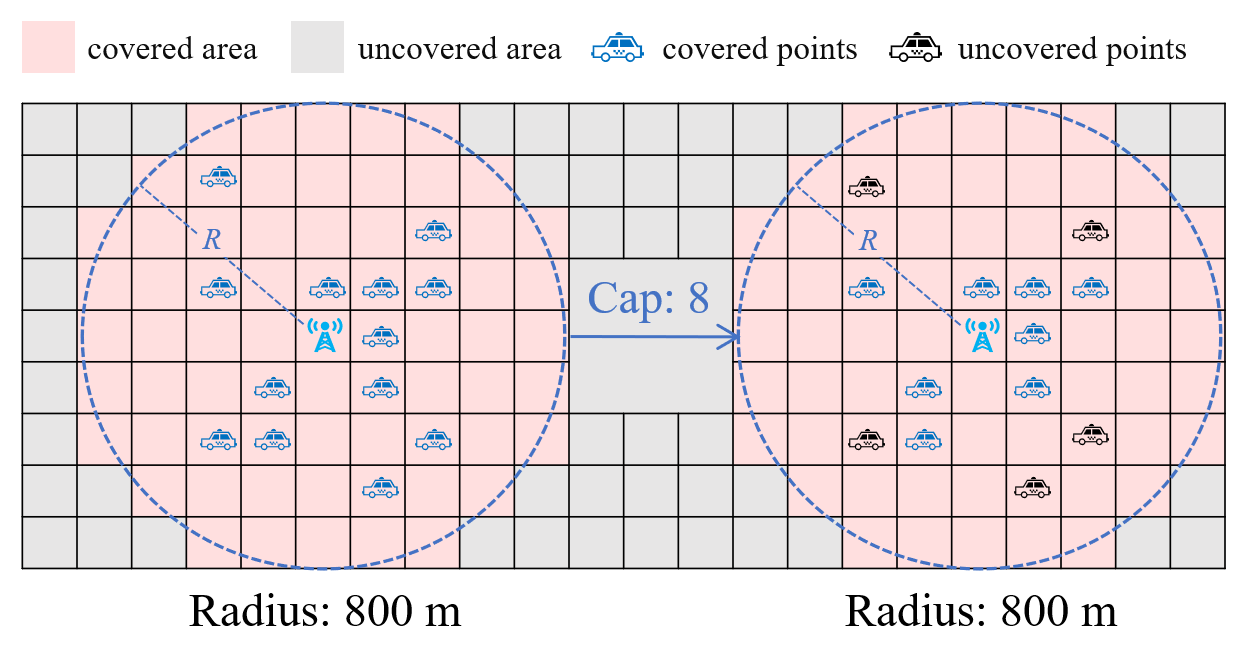}
  \caption{Different Coverage of Potential Edge Servers.}
  \label{fig:diagram} 
\end{figure}

\textbf{Area Coverage}: The ratio of the area (i.e., number of grids) covered by the edge servers to the total area of the study region for a given time duration.  
In Fig.~\ref{fig:diagram}, the light red grids are covered, while the grey ones are uncovered. 
A higher area coverage indicates a greater probability that a user can be served at a random location and time point, which is a prerequisite for deploying edge computing services.  

\textbf{Point Coverage}: The ratio of the number of covered demand points to the total number of demand points, assuming unlimited capacity of edge servers. 
 A demand point is a computation request generated by a taxi at a specific location on the map, corresponding to a GPS record in the Shanghai taxi dataset. In the left part of Fig.~\ref{fig:diagram}, all the points are covered by the server with unlimited capacity.  
 This metric accounts for user dynamics and evaluates the geographic distribution of the servers to meet the spatiotemporally varying user demands. 

\textbf{Point Coverage with Cap}: This metric is similar 
to the previous metric. The difference is that Point Coverage with Cap enforces limited computation capacity of each server, while Point Coverage does not. If the number of demand points overwhelms the server capacity, they will not be covered even if they are within the service range of the server. 
In the right part of Fig.~\ref{fig:diagram}, the computation capacity of the server is $8$ per time slot, so the $8$ blue points are covered while the $5$ black ones are not.  
This metric better reflects the actual ability of edge servers to meet the dynamic user demands. 



\subsubsection{Coverage of Base Stations}
\begin{figure*}
    \centering
    \begin{subfigure}{0.32\textwidth}
        \centering
        \includegraphics[width=\linewidth]{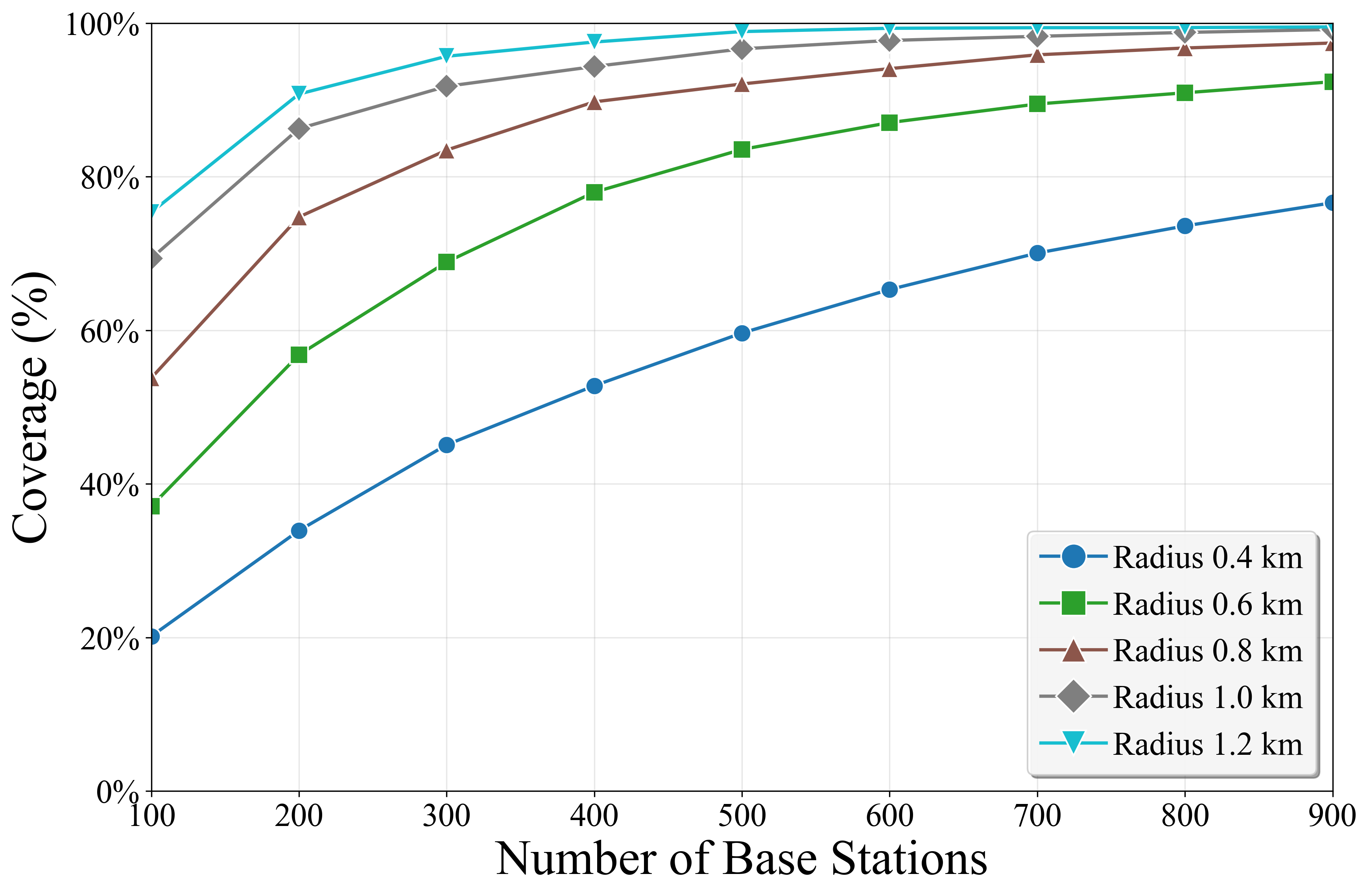}
        \caption{Area Coverage}
        \label{fig:station_area}
    \end{subfigure}
    \hfill
    \begin{subfigure}{0.32\textwidth}
        \centering
        \includegraphics[width=\linewidth]{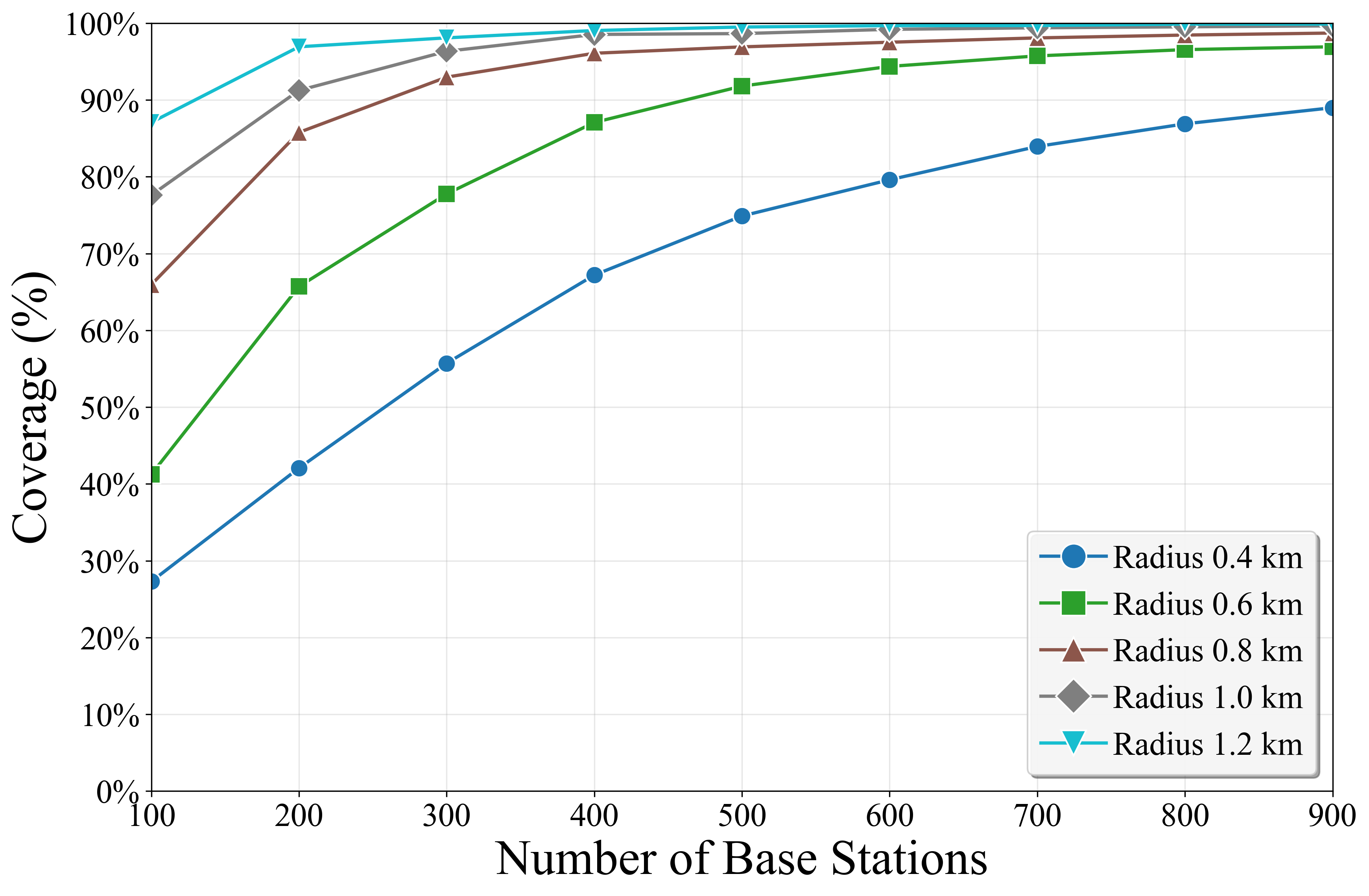}
        \caption{Point Coverage}
        \label{fig:station_point}
    \end{subfigure}
    \hfill
    \begin{subfigure}{0.32\textwidth}
        \centering
        \includegraphics[width=\linewidth]{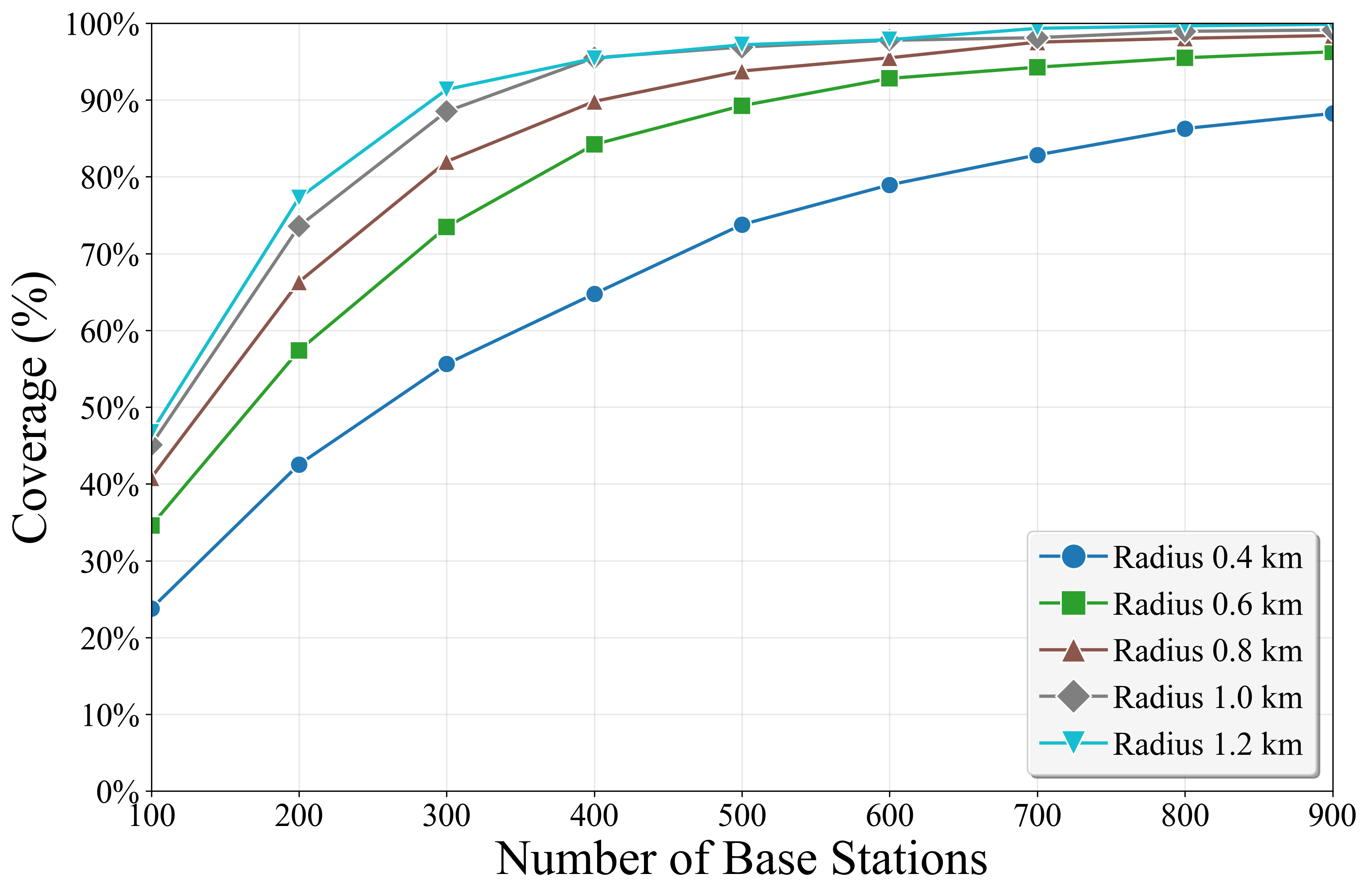}
        \caption{Point Coverage with Cap}
        \label{fig:station_point_with_cap}
    \end{subfigure}
    \caption{Coverage of Base Stations.}
    \label{fig:station_three}
\end{figure*}

The coverage of the base stations on the demand points on April 1st in $2018$ is shown in Fig.~\ref{fig:station_three}. We select different portions of base stations randomly to deploy edge servers, and also vary their service radius. The number of randomly selected base stations is varying from $100$ to $900$,  and the radius is varying from $0.4$ km to $1.2$ km.

Fig.~\ref{fig:station_area} illustrates the area coverage of the base stations, showing that
the existing base station layout can cover most of the area. For instance, 
with a service radius of $0.8$ km, $500$ base stations yield over $90\%$ area coverage. 
Fig.~\ref{fig:station_point} shows the point coverage, and the result also indicates a near-perfect coverage when the server computation capacity limit is ignored.
However, Fig.~\ref{fig:station_point_with_cap} shows much worse results when we adopt a more realistic setting by limiting the server capacity to $50$ requests per time slot. 
The coverage decreases remarkably for all radii, which indicates that the capacity is in fact the primary bottleneck rather than the radius. It is worth noting that deploying edge servers  with sufficient capacity to cover all demands all the time is irrational, as such capacity during normal periods may be wasted.

\subsubsection{Coverage of Buses}
\begin{figure*}
    \centering
    \begin{subfigure}{0.32\textwidth}
        \centering
        \includegraphics[width=\linewidth]{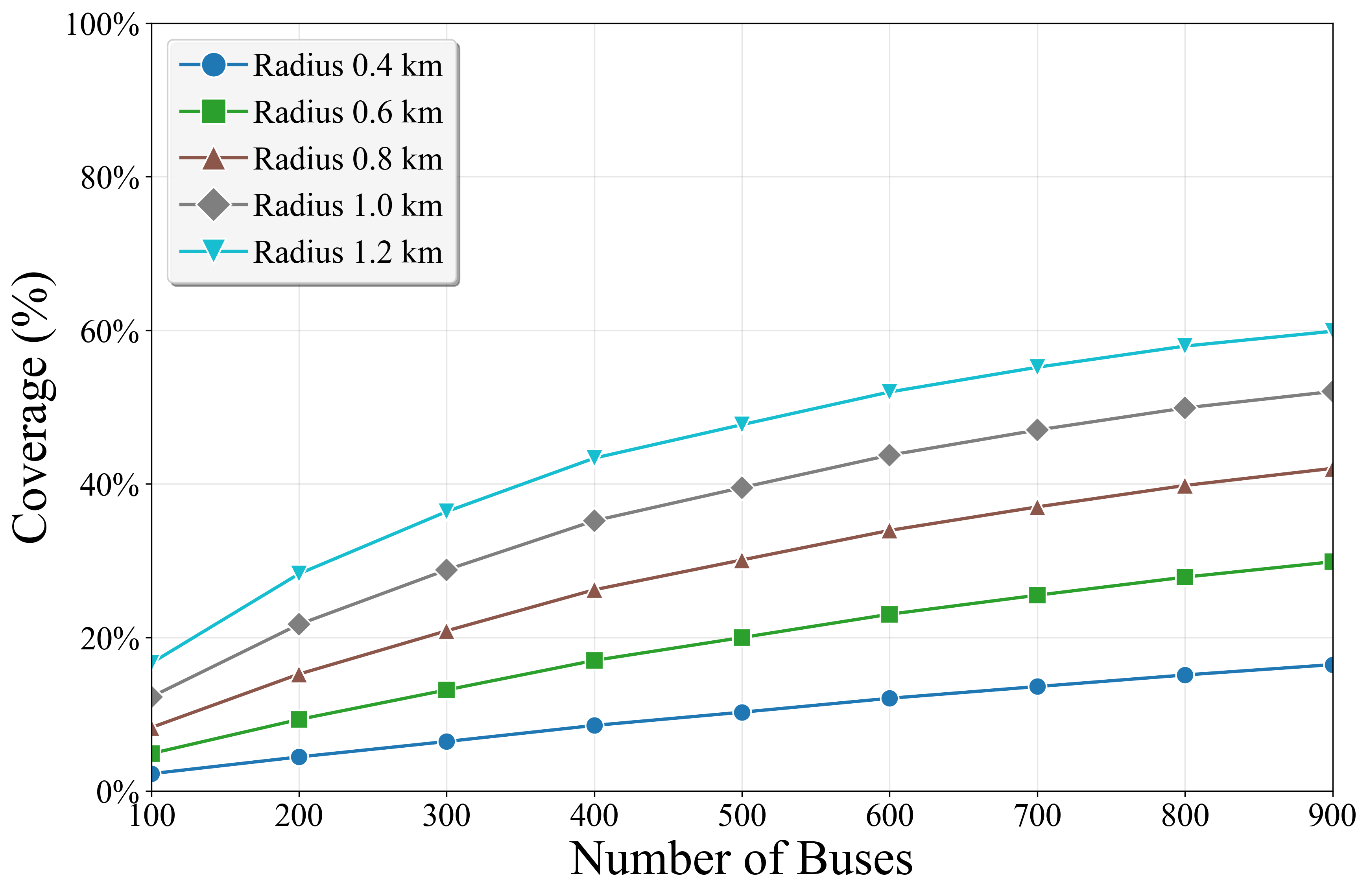}
        \caption{Area Coverage}
        \label{fig:bus_area}
    \end{subfigure}
    \hfill
    \begin{subfigure}{0.32\textwidth}
        \centering
        \includegraphics[width=\linewidth]{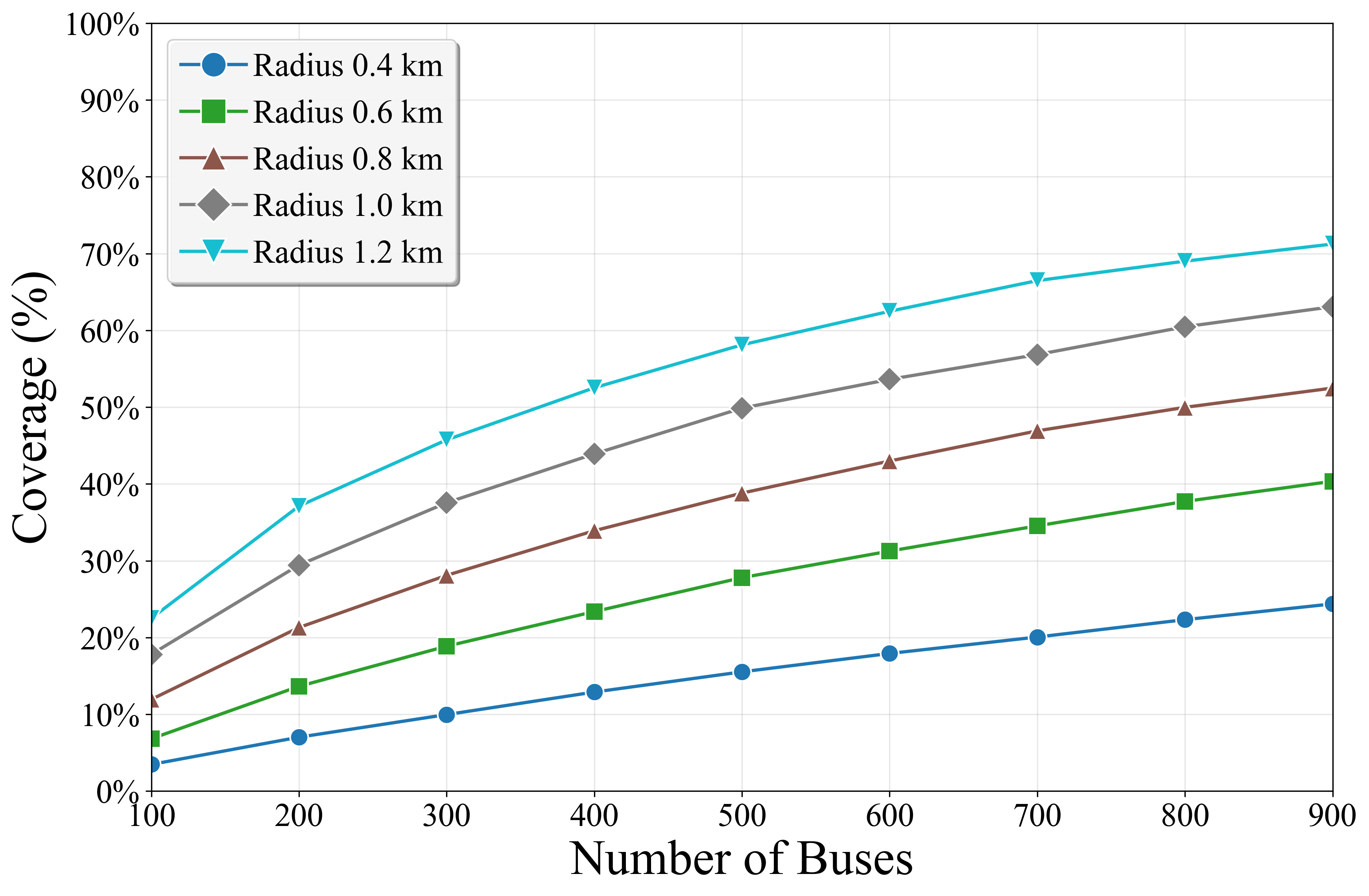}
        \caption{Point Coverage}
        \label{fig:bus_point}
    \end{subfigure}
    \hfill
    \begin{subfigure}{0.32\textwidth}
        \centering
        \includegraphics[width=\linewidth]{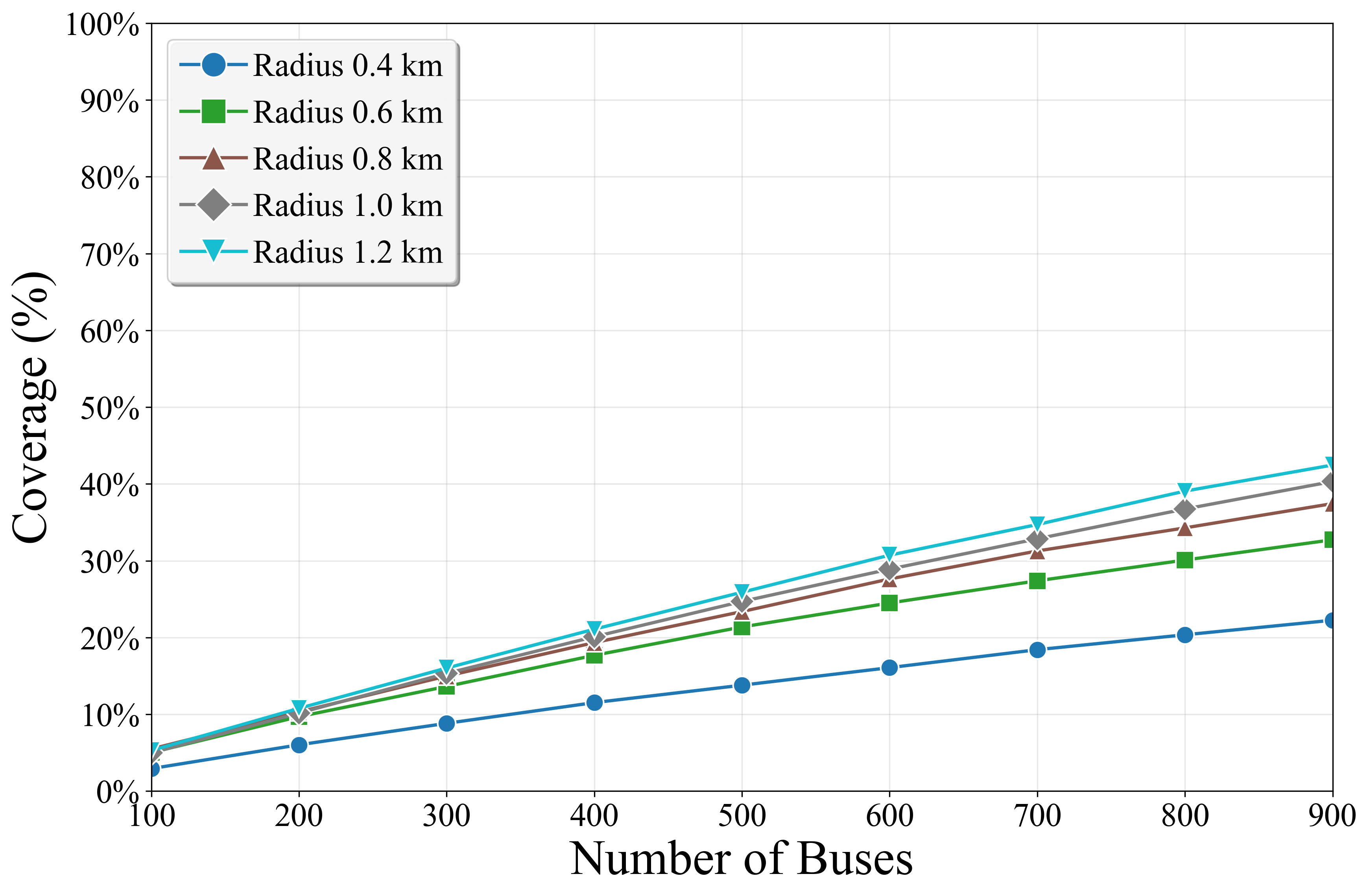}
        \caption{Point Coverage with Cap}
        \label{fig:bus_point_with_cap}
    \end{subfigure}
    \caption{Coverage of Buses.}
    \label{fig:bus_three}
\end{figure*}

We then evaluate the coverage of the buses if they are employed as mobile edge servers in Fig~\ref{fig:bus_three},  following similar settings to the base stations.  

The results show that buses also demonstrate desirable performance  
in terms of the three coverage metrics. 
The area and point coverage of buses, although smaller than base stations, can also reach over $50\%$ for certain server percentages and radii as shown in Fig.~\ref{fig:bus_area} and Fig.~\ref{fig:bus_point}.   
Even under more realistic scenarios shown in Fig.~\ref{fig:bus_point_with_cap}, the buses still cover a considerable percentage of the user demands. 
With a service radius of $0.8$ km and $900$ random buses equipped with edge servers, nearly $40\%$ of the demand points can be covered.  
This indicates that the buses always go through high-traffic zones, which have a great potential in acting as edge servers and compensating for the server capacity deficit of base station-based edge servers. 

\section{Selection Strategy of Bus-Mounted  Servers}\label{sec:selection}
In this section, we present our system model and algorithm design of the bus selection strategy. We aim at deliberately selecting a limited number of buses that offer the best coverage, rather than selecting them randomly in the previous section.

\subsection{System Model}
 We model the city using a grid system with $G$ grids:  $\mathcal{G} = \{1, 2, \ldots, G\}$,  
 and $T$ discrete time slots $\mathcal{T} = \{1, 2, \ldots, T\}$. The duration of each time slot is $\tau$, with a typical value of $\tau = 30$ seconds, which captures the dynamic mobility of both buses (edge servers) and taxis (end users). 
 $\mathrm{dist}_{i,j}$ is the Euclidean distance between grid $i$ and $j$. 
 The spatiotemporal changing demand points are modeled using a series of $\mathcal P(t)=\{l_i^{\rm dp}(t):i=1,...,N(t),l_i^{\rm dp}(t)\in\mathcal{G}\}$, each meaning the multiset of demand points (grid IDs as the locations) in time slot $t$, i.e., multiple demand points may appear at the same grid. $N(t)$ is the total number of demand points in that time slot.  Note that such demand can be the total demand of end users, or the remaining demand after subtracting the demand offloaded to base stations. 
Each bus-mounted edge server has a fixed computation capacity $C^{\rm bus}$, i.e., the maximum number of demand points it can serve per time slot, and a fixed service radius $R$. The set of all the bus trajectories is $\mathcal{R}=\{\,r_1,r_2,\ldots,r_M\}$, where each bus trajectory $r_j=\{\,l^{\rm bus}_j(1),l^{\rm bus}_j(2),\ldots,l^{\rm bus}_j(T)\}$ contains a series of locations (grid IDs) in different time slots, i.e., $l^{\rm bus}_j(t)=g, t\in\mathcal{T},g\in\mathcal{G}$. 

The selection of buses is represented using a binary vector
$\mathbf{s}=[s_1,s_2,\ldots,s_M]$, where $s_i=1$ means bus $i$ is selected to carry an edge server and offer computation offloading services, and $s_i=0$ otherwise. Due to budget limit, the problem is to find $K$ buses out of $M$ to maximize the coverage of the demand points for the entire time duration. Therefore,
\begin{equation}
\sum_{i=1}^Ms_i = K.
\end{equation}

The coverage of demand points is modeled using $\theta_{i,j}(t)$, where $\theta_{i,j}(t)=1$ means demand point $l_i^{\rm dp}(t)$ is assigned to bus $j$ in time slot $t$, and $\theta_{i,j}(t)=0$ otherwise. Thus the number of demand points covered by bus $j$ in time slot $t$ is
\begin{equation}
c_j(t)=s_j\sum_{i=1}^{N(t)}\theta_{i,j}(t).
\end{equation}
Such coverage must be under the server capacity constraint
\begin{equation}
c_j(t)\leq C^{\rm bus},
\end{equation}
and within the service radius: 
\begin{equation}
{\rm if}\;{\rm dist}_{l_i^{\rm dp}(t),l^{\rm bus}_j(t)}>R: \theta_{i,j}(t)=0.
\end{equation}

Therefore, the total number of covered demand points of all the buses throughout all the time slots is 
\begin{equation}
CDP = \sum_{t=1}^{T}\sum_{j=1}^{M}c_j(t), 
\end{equation}
and the optimization problem is to find $K$ buses that maximize the number of covered demand points, formulated as
\begin{align}
\textbf{Maximize:\;}& CDP = \sum_{t=1}^{T}\sum_{j=1}^{M}c_j(t), \label{eq:obj} \\
\textbf{s.t.} \quad 
& \sum_{i=1}^Ms_i = K, \\
& \forall j,t:c_j(t)=s_j\sum_{i=1}^{N(t)}\theta_{i,j}(t),\\
& \forall j,t:c_j(t)\leq C^{\rm bus}, \\
&\forall i,j,t:{\rm if}\;{\rm dist}_{l_i^{\rm dp}(t),l^{\rm bus}_j(t)}>R: \theta_{i,j}(t)=0,\\
& \forall i,j,t:s_i,\theta_{i,j}(t)\in\{0,1\},
\end{align}
where $s_i$-s and $\theta_{i,j}(t)$-s are binary unknown variables. 
Obviously, this problem is an integer quadratic programming problem, which is NP-hard, so we design a simple greedy heuristic algorithm to solve it. 


\subsection{Greedy Heuristic Algorithm}
We propose a simple greedy heuristic algorithm to solve Problem~\eqref{eq:obj}. 
The algorithm iteratively selects the bus with the maximum coverage of demand points on all the time slots. 

For each iteration, we first 
calculate the total number of covered demand points $C_i$ for each bus $i$ independently throughout all the time slots, assuming a bus does not consider the choice of other buses. The assignment of the demand points to buses follow a simple closest-first principle, i.e., a bus serves the closet points within its service range until its computation capacity is fully saturated. Next, the bus covering the greatest number of demand points is selected and added to the result set, and all the covered demand points by this bus are removed from the map. This process repeats at most $K$ times to select $K$ buses, or until all the demand points are covered. 
The algorithm is detailed in Alg.~\ref{alg:bus_route_selection}.

\begin{algorithm}
\caption{Greedy Selection of Buses}
\label{alg:bus_route_selection}
\begin{algorithmic}[1]
\REQUIRE 
    $\mathcal G$: set of grids;
    $\mathcal{R}$: set of buses; 
    $\{\mathcal{P}(t)\}$: set of demand points; 
    $R$: bus service radius; 
    $C^{\rm bus}$: bus computation capacity; 
    $K$: number of selected buses; 
\ENSURE 
    $\mathcal{S}$: set of $K$ selected buses;
    \STATE Initialize the  set of selected buses: $\mathcal{S} \leftarrow \emptyset$;
    \FOR{each $iter \in 1\rightarrow K$}
        \FOR{each bus $i \in 1\rightarrow M$}
             \FOR{each time slot $t \in \mathcal{T}$}
                \STATE Initialize the set of covered demand points by bus $i$ in time slot $t$:  $\mathcal{P}_i(t)\leftarrow \emptyset$;
                \STATE Find  up to $C^{\rm bus}$ closest demand points to bus $i$ within its service radius $R$ and add them to $\mathcal{P}_i(t)$; 
                
            \ENDFOR 
            \STATE Construct the set of covered demand points of bus $i$ over all time slots: $\mathcal{P}_{i}\leftarrow \{\mathcal{P}_i(1),\mathcal{P}_i(2),\cdots\mathcal{P}_i(T)\}$;
            \STATE Calculate the number of covered demand points by bus $i$: $C_{i}\leftarrow \sum_{t=1}^T |{P}_i(t)|$;
        \ENDFOR
        \STATE Find the bus $x$ with maximum  number of covered demand points: $x={\rm argmax}_{i=1}^M C_i$;
        \STATE Add bus $x$ to the selection set: $\mathcal{S}\leftarrow\mathcal{S}+x$;
        \STATE Remove bus $x$ from the bus set: $\mathcal{R}\leftarrow\mathcal{R}-x$;
         \STATE Remove the demand points covered by bus $x$: $\{\mathcal{P}(t)\}\leftarrow\{\mathcal{P}(1)-\mathcal{P}_x(1),\mathcal{P}(2)-\mathcal{P}_x(2),\cdots,\mathcal{P}(T)-\mathcal{P}_x(T)\}$;
        \IF {$\forall t:\mathcal{P}(t)=\emptyset$} 
            \STATE break;
        \ENDIF
    \ENDFOR
\RETURN $\mathcal{S}$;
\end{algorithmic}
\end{algorithm}

\section{Performance Evaluation}\label{sec:eval}
In this section, we evaluate the performance of our greedy heuristic bus selection algorithm. 

\subsection{Experiment Setup}
We conduct simulations using the bus/taxi GPS records and base station locations from the Traces in Shanghai. Each grid is $200\, {\rm m}\times 200\, {\rm m}$ and each time slot lasts $30$ seconds. To challenge our bus-based edge computing strategy most, we assume there are no edge servers deployed at base stations at all. 
The performance of our proposed greedy bus selection algorithm is compared with the following benchmark algorithms.
\begin{itemize}
    \item 
    \textit{Random Selection}: This method randomly selects the predetermined number of buses. 
    \item 
    \textit{Cluster-Based Selection}: This algorithm first clusters the demand points, and then ranks the buses based on their coverage of major demand clusters. The buses with strongest coverage capability will be selected. 
\end{itemize}

\subsection{Performance Results}
\begin{figure}[h!] 
  \centering
  \includegraphics[width=0.8\linewidth]{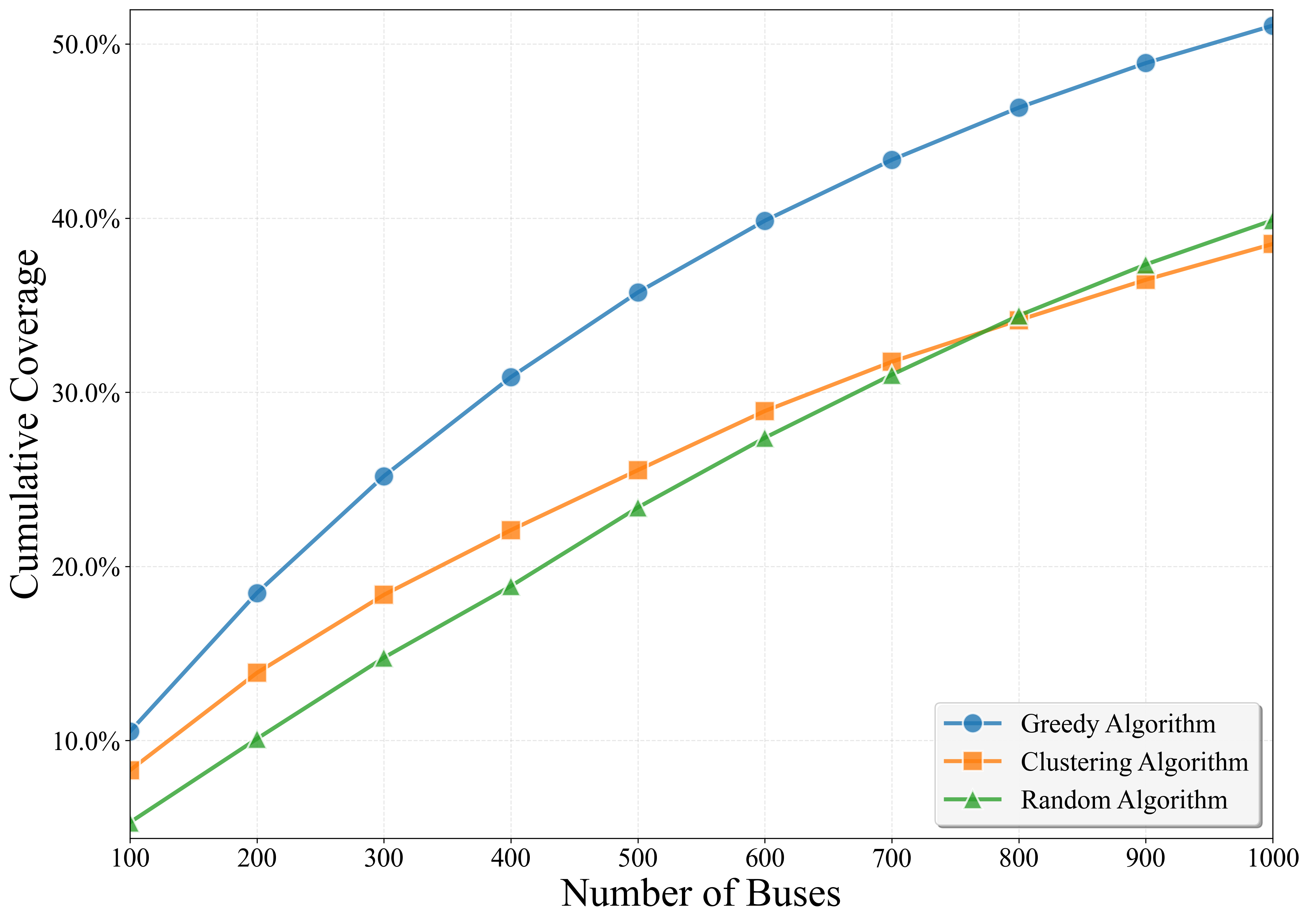}
  \caption{Coverage with Cap under Different Algorithms.}
  \label{fig:dif_algorithm} 
\end{figure}

Fig.~\ref{fig:dif_algorithm} shows the coverage ratio of all the selection algorithms, where the service radius of bus-mounted edge servers is set to $R=0.8$ km, and computation capacity is $C^{\rm bus}=50$ tasks per time slot. We observe that our proposed greedy heuristic algorithm consistently outperforms other algorithms. 
Specifically, the greedy algorithm with 500 buses demonstrate a coverage ratio of $35.7\%$ in comparison with $23.3\%$ of the random algorithm, a $50\%+$ performance improvement. With $1,000$ buses, the coverage ratio further improves to $51\%$.  
This shows that our greedy algorithm greatly improves the bus selection performance and unleashes their potential of carrying edge servers.


\begin{figure}[h!] 
  \centering
  \includegraphics[width=0.8\linewidth]{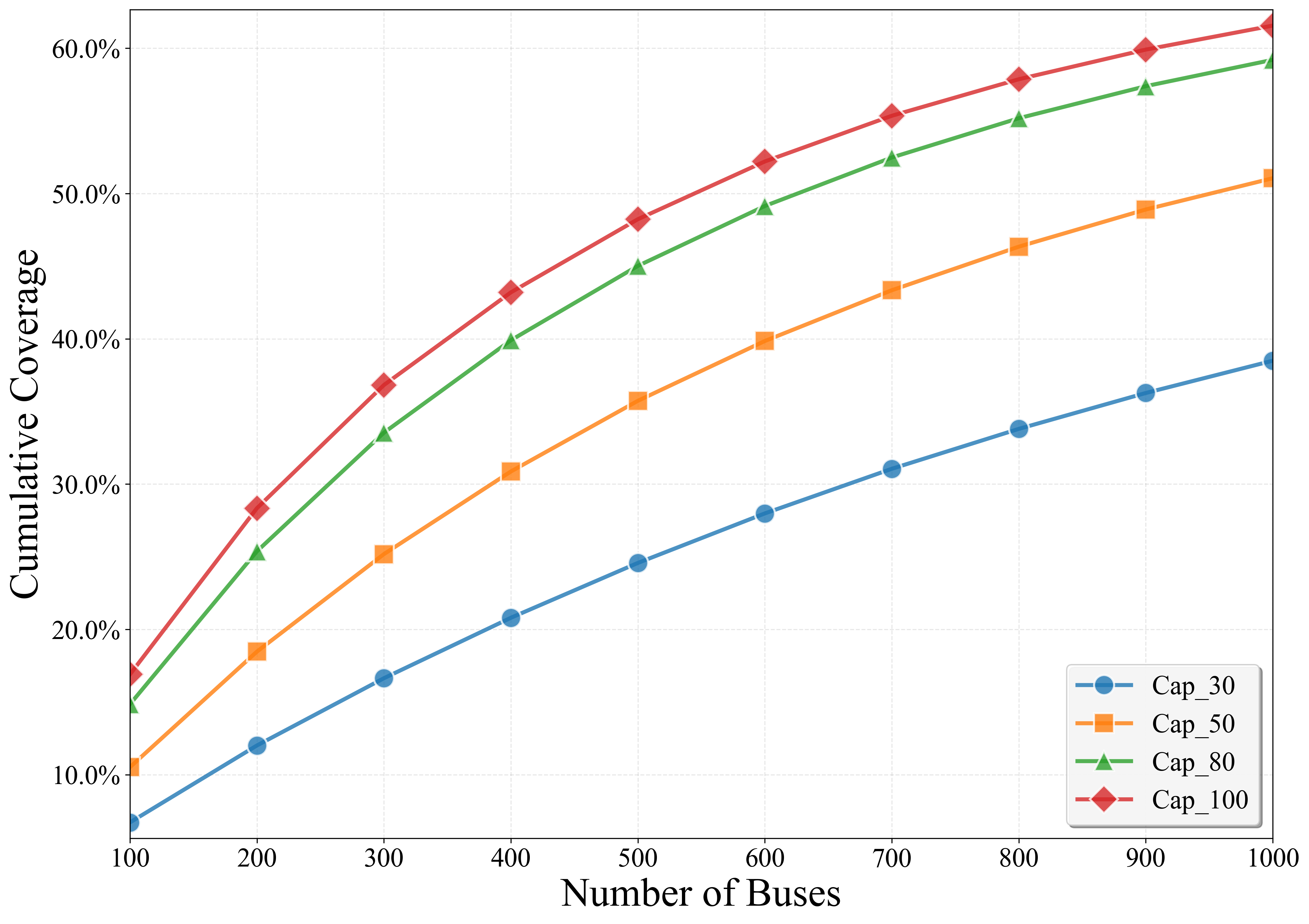}
  \caption{Coverage under Different Caps.} 
  \label{fig:dif_cap} 
\end{figure}

Fig.~\ref{fig:dif_cap} shows the coverage ratio of different numbers of selected buses with varying computation capacity. The figure shows that the coverage ratio increases monotonically with the number of selected buses. When the number of buses is $1,000$, the coverage ratio can reach nearly $40\%$ for a computation capacity $C^{\rm bus}=30$, and over $60\%$ for $C^{\rm bus}=100$.  
By increasing the number of buses and computation capacity, our method can further approach the global optimum, demonstrating that it can achieve efficient coverage.


\begin{figure*}[!htp]
    \centering
    \begin{subfigure}{0.32\textwidth}
        \centering
        \includegraphics[width=\linewidth]{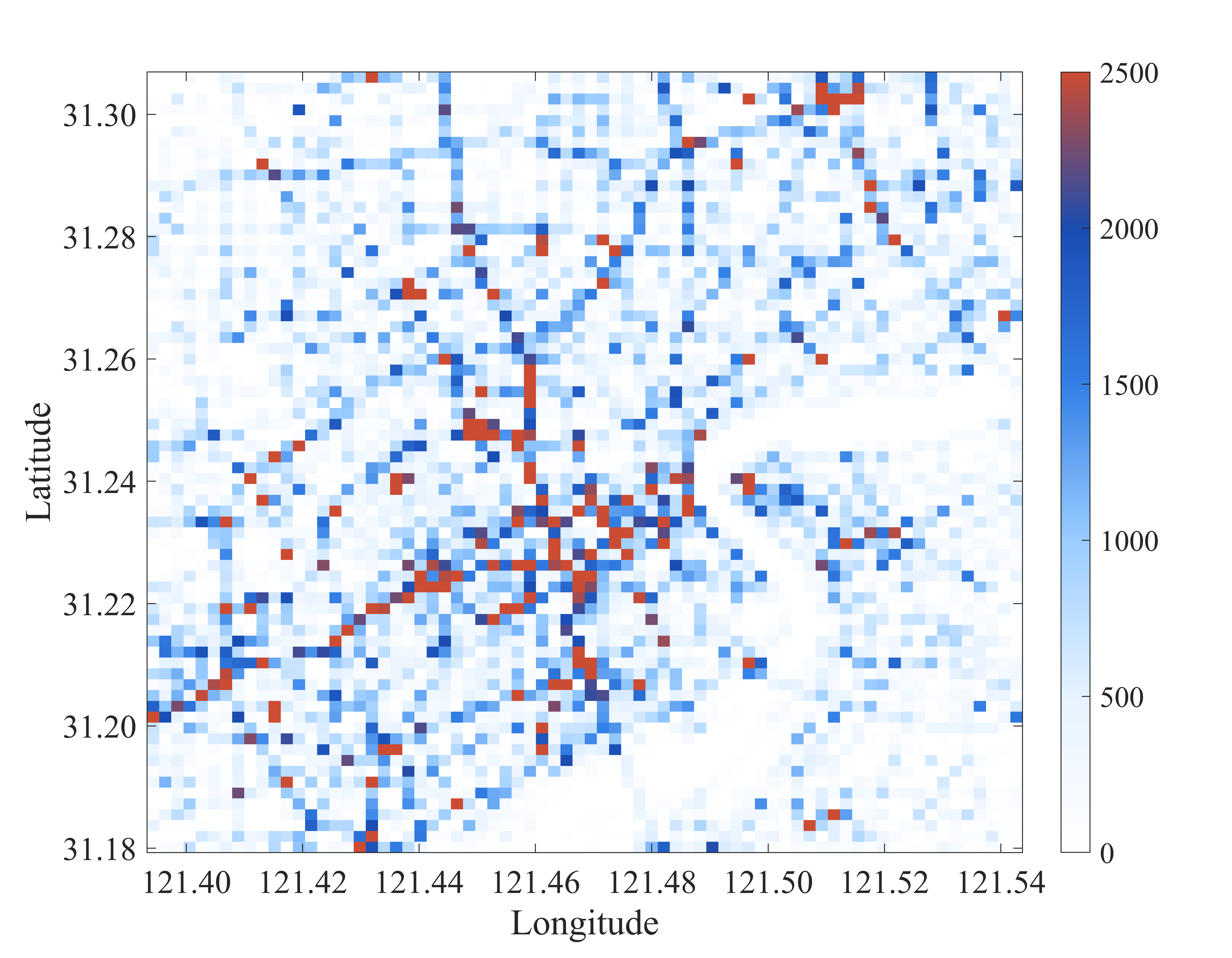}
        \caption{Demands without Edge Computing}
        \label{fig:original_state}
    \end{subfigure}
    \begin{subfigure}{0.32\textwidth}
        \centering
        \includegraphics[width=\linewidth]{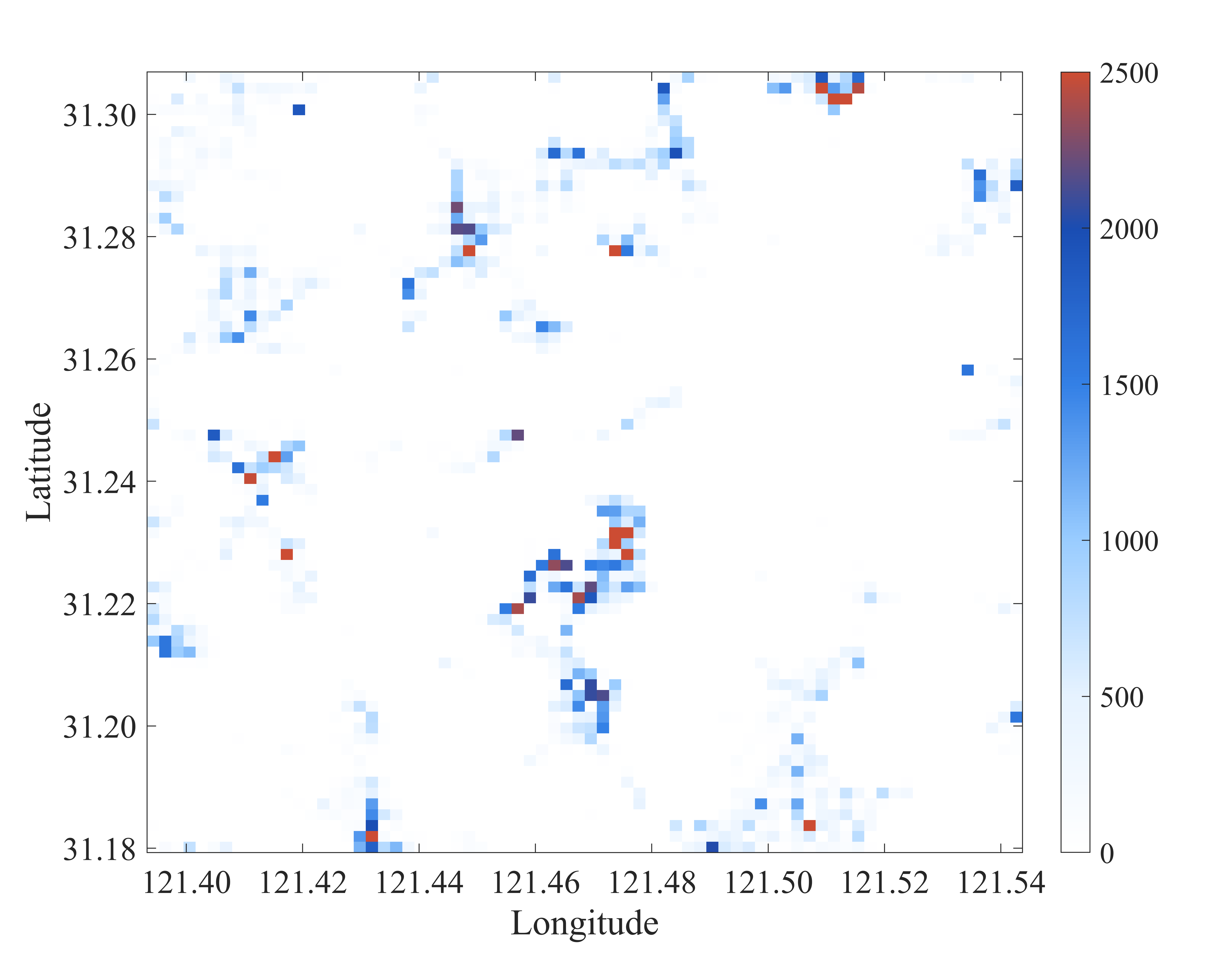}
        \caption{After Processed by BSs}
        \label{fig:300_200_BS}
    \end{subfigure}
    \begin{subfigure}{0.32\textwidth}
        \centering
        \includegraphics[width=\linewidth]{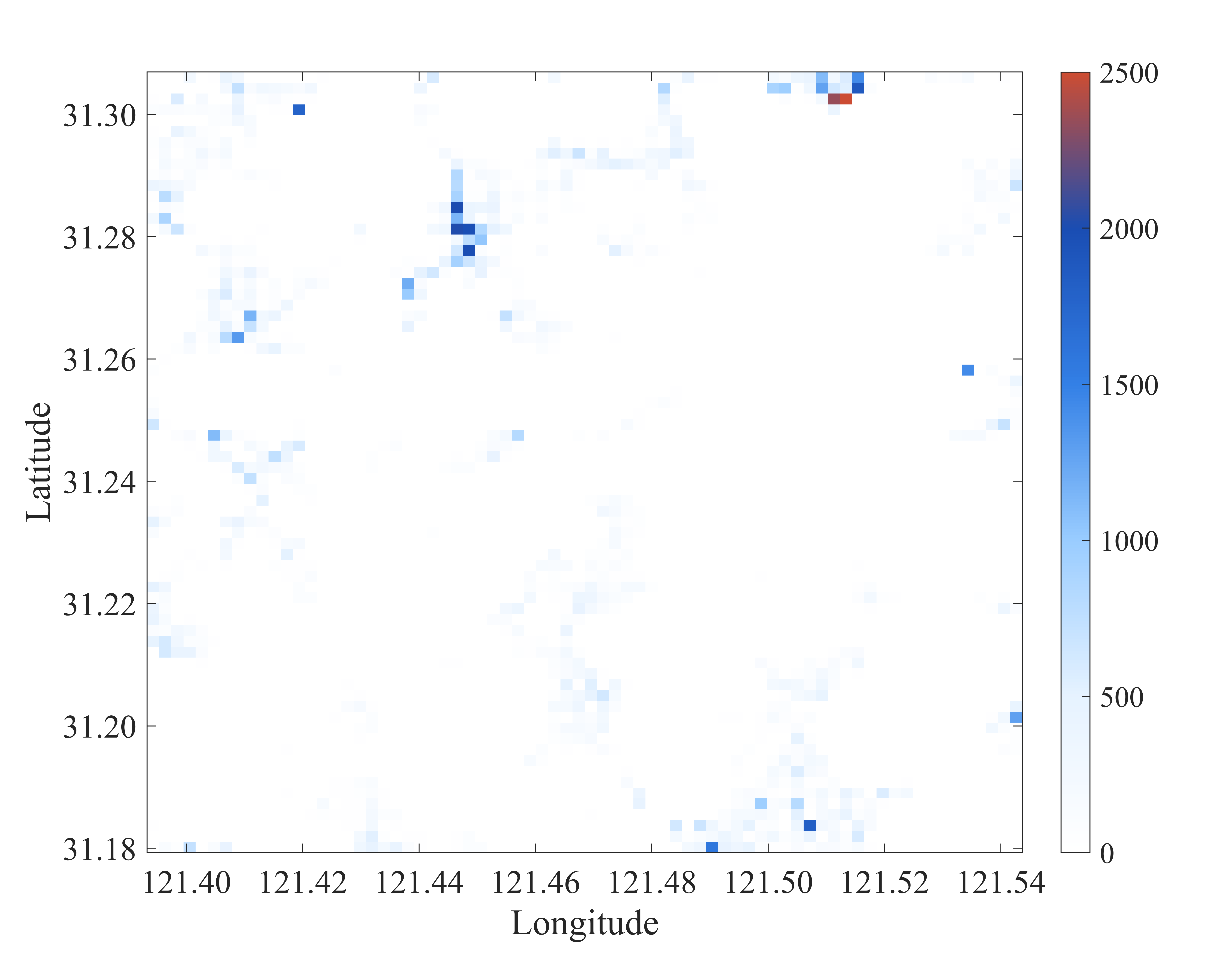}
        \caption{After Processed by BSs and Buses}
        \label{fig:300_200_BSandBus}
    \end{subfigure}
    \caption{Demands without and with Base Station/Bus-based Edge Servers.}
    \label{fig:evening_peak_load}
\end{figure*}

Fig.~\ref{fig:evening_peak_load} shows a comparison of the demand volume covered by base stations and buses 
during the evening peak hour from 18:00 to 20:00, to show the effect of deploying both base station and bus-based edge servers.  Fig.~\ref{fig:original_state} shows the demands of the end users without any edge servers.   
Fig.~\ref{fig:300_200_BS} shows the remaining demands after deploying $300$ random base stations, and Fig.~\ref{fig:300_200_BSandBus} shows the result after introducing additional $200$ heuristically selected buses on that basis.   Both the base stations and buses have a service radius of $0.8$ km and a capacity of $50$. The results reveal that while base stations handle most of the demands, some downtown areas still exhibit high load (the grids with red color), yet the buses effectively eliminate these remaining demand hotspots.  This indicates that bus-mounted edge servers can cooperate with the base stations well, effectively handle the bursty traffic overwhelming the computation capacity of base stations during rush hours, and thereby improve the system computation elasticity and overall quality of service. 


\section{Conclusion and Future Work}\label{sec:conclu}
This paper studied the feasibility of bus-mounted edge servers in urban areas. First, the coverage of the buses and base stations were investigated through a trace study based on the Shanghai bus/taxi/Telecom datasets. Then, we built a mathematical model and designed a simple greedy heuristic algorithm for optimizing the bus selection that maximized the coverage of demand points. 
We performed trace-driven simulations to verify the performance of the proposed bus selection algorithm, and the results showed that our approach effectively improved the spatiotemporal coverage under realistic constraints such as server capacity and purchase quantity.


In our future work, we plan to extend our problem to multi-objective optimization, e.g., incorporating more objectives such as energy consumption. 
We will also work on fine-grained management of bus-mounted edge servers, e.g., where and when to work to improve system QoS, or to turn dominant to save energy. 

\section*{Acknowledgments}
This work was partially supported by the National Natural Science Foundation of China (No. 62371142, No. 62273107 and No. 62473381), the Guangdong Basic and Applied Basic Research Foundation (No. 2024A1515010404), and the Key Research and Development Program of Hunan Province (No. 2025JK2068).

\end{document}